\documentclass[12pt, nofootinbib]{revtex4-2}
\usepackage{hyperref}
\usepackage{appendix}
\usepackage{subfigure}
\usepackage{graphicx}% Include figure files
\usepackage{dcolumn}% Align table columns on decimal point
\usepackage{bm}
\usepackage{amsmath}
\usepackage{amssymb}
\usepackage{color}

\def\bea {\begin{eqnarray}}
\def\eea {\end{eqnarray}}

\def\di {\partial}

\begin{document}
\title{Entanglement production through a cosmological bounce}
  
\author{Viqar Husain} \email{vhusain@unb.ca}
\affiliation{Department of Mathematics and Statistics, University of New Brunswick, Fredericton, NB Canada E3B 5A3 }

\author{Irfan Javed}  \email{i.javed@unb.ca}
 \affiliation{Department of Mathematics and Statistics, University of New Brunswick, Fredericton, NB Canada E3B 5A3 }
 
 \author{Sanjeev S.\ Seahra} \email{sseahra@unb.ca}
 \affiliation{Department of Mathematics and Statistics, University of New Brunswick, Fredericton, NB Canada E3B 5A3 }

\author{Nomaan X}  \email{nomaan.math@unb.ca}
 \affiliation{Department of Mathematics and Statistics, University of New Brunswick, Fredericton, NB Canada E3B 5A3 }

%\date{\today} 
 
\begin{abstract}
\vskip 1cm

In quantum cosmology, it is expected that the Big Bang singularity is resolved and the universe undergoes a bounce. We find that for Gaussian initial states, matter-gravity entanglement entropy rises rapidly during the bounce, declines, and then approaches a steady-state value following the bounce. These observations suggest that matter-gravity entanglement is a feature of the macroscopic universe and that there is no Second Law of entanglement entropy.

\vskip 2.5cm

\begin{center}
\noindent{This essay received an honorable mention in the 2024 \\ Gravity Research Foundation competition.}
%Essay written for the Gravity Research Foundation\\ 2024 Awards for Essays on Gravitation}
\end{center}
\vskip 7.0cm
\noindent Corresponding author: Viqar Husain

\end{abstract}

\maketitle
\eject

 Classical gravity is described by coupled equations for the degrees of freedom of gravity and matter. In the conventional application of quantization to gravity, the Hilbert space of states therefore takes the form of a bipartite system---the tensor product of gravity and matter Hilbert spaces: ${\cal H}_{QG} = {\cal H}_{G}\otimes {\cal H}_{M}$. The matter Hilbert space is in general also a tensor product of the spaces for various species of matter. This tensor product structure could lead to matter-matter or matter-gravity entanglement. Either is a potential observable for quantum gravity \cite{Bose:2017nin,Marletto:2017kzi,Krisnanda_2020, kay2018matter}, one which may play a role in understanding the emergence of quantum fields on curved spacetime from a theory of quantum gravity.

 In the absence of a complete and accepted theory of quantum gravity in four spacetime dimensions \cite{Loll:2022ibq}, attempts to obtain insights have focused on model systems \cite{Kuchar:1989tj,BarberoG:2010oga}, the equivalent of Bohr atoms. Among these are cosmological models; the main result that has emerged from a variety of studies is the avoidance of the Big Bang singularity, a feature of this effect being that the universe undergoes a bounce after shrinking to a Planck scale volume. 
 
 %\sss{Possibly a side comment for a larger paper: is the volume at the bounce in this model at the Planck scale?  It would seem to depend on the assumed momentum of the scalar field I would guess.}

 An interesting question is what happens to the entanglement between the states of matter and gravity as the universe undergoes a bounce. Intuition suggests that such entanglement should be large in the deeply quantum regime due to the strong coupling between matter and gravity, and that it should decline as the universe expands and dynamically evolves into the ``semiclassical" era of classical gravity and quantum matter. This is the question we examine in this essay. 
 
 Parallels of this question have been explored in particle physics, where it has been argued that interactions produce entanglement between the system and the apparatus in QCD \cite{Kovner_2019}; in models of spins coupled to an oscillator, where the entanglement entropy can oscillate \cite{Husain:2022kaz}; and in systems of coupled oscillators, where the entropy rises monotonically and saturates \cite{Husain:2023jaq} to a value that depends on the system's total energy, following the law $S = 2/3 \ln E$.

We study this question in quantum cosmology with three degrees of freedom, the universe's volume $v$, a dust field $T$, and a scalar field $\varphi$. The canonical description of the model has these configuration degrees of freedom and their conjugate momenta $p_v$, $p_T$, and $p_\varphi$, with evolution given by the Hamiltonian constraint \cite{Husain:2019nym,Gielen:2020abd}. The dust field provides a useful time variable, leading to a physical ``relational" Hamiltonian for the coupled dynamics of $v$ and $\varphi$. The Hamiltonian (in units $G=c=\hbar=1$) is 
\bea 
 H = p_v^2 - \frac{1}{v^2}\ p_\varphi^2;
 \label{H}
\eea
there are immediate generalizations with a scalar potential, cosmological constant, and nonzero curvature, but we consider here the simplest case, which is sufficient to address the question of entanglement through the bounce.

This physical Hamiltonian is typical of the form that arises from the Hamiltonian constraint of general relativity---it is unlike that for usual systems that contain a positive kinetic term for each particle species and additive interactions; in the Hamiltonian (\ref{H}), the term $\displaystyle -p_\varphi^2/v^2$ comes from the universal coupling of gravity to matter.

The corresponding time-dependent Schrodinger equation for the wave function $\psi(t,v,\varphi)$ is
\begin{equation}
i \frac{\di\psi}{\di t} = - \frac{\di^{2} \psi}{\di v^{2} }  - \frac{1}{v^{2}} \left(\mu - \frac{\di^{2}}{\di\varphi^{2}} \right) \psi, \quad v \in (0,\infty),
\label{tdse}
\end{equation}
where $t$ is the time in the dust gauge $T=t$ \cite{Husain:2011tk}, and $\mu =1/4$ and $\mu=0$ realize two representations of the Hamiltonian operator used respectively in \cite{Gielen:2020abd} and \cite{Husain:2019nym}. The Schrodinger equation (\ref{tdse}) may be solved numerically as in \cite{Husain:2019nym} and also analytically  by separation of variables as in \cite{Gielen:2020abd} in terms of Bessel functions. In this work, we find it convenient to solve the equation numerically. To do so, we need to specify boundary conditions on the wave function.  In order to calculate entanglement entropy through the bounce, it is useful to assume periodic boundary condition in the field direction $\varphi$; that is,
\begin{equation}
    \psi(t,v,\varphi+R) = \psi(t,v,\varphi),
\end{equation}
where $R$ is the period (which can be taken to be as large as desired). In order to have a well-defined numerical scheme, we also assume Dirichlet boundary conditions in the $v$ direction:
\begin{equation}
    \lim_{v\rightarrow 0}\psi(t,v,\varphi)= \lim_{v\rightarrow \infty}\psi(t,v,\varphi)=0.
\end{equation}
These boundary conditions ensure that the inner product between two solutions $\psi_1$ and $\psi_2$ of the Schrodinger equation is conserved:
\begin{equation}
    \frac{d}{dt} \iint dv \, d\varphi \, \psi_1^* \psi_2 = 0.
\end{equation}

\textcolor{black}{Due to the periodicity in the field direction, solutions of the Schrodinger equation could be expressed as
\begin{equation}
	\psi(t,v,\varphi) = \frac{1}{\sqrt{R}}\sum_{n=-\infty}^\infty c_{n} \exp \left( i \frac{2n\pi\varphi}{R} \right) \xi_{n}(t,v),
 \label{mode}
\end{equation}
where $\xi_{n}(t,v)$ satisfies
\begin{equation}
i \frac{\di\xi_{n}}{\di t} = - \left( \frac{\di^{2} }{\di v^{2} }  + \frac{\alpha_n}{v^{2}}\right) \, \xi_{n}, \quad \alpha_{n} = \mu +\frac{4\pi^{2}n^{2}}{R^{2}}.
\label{modeq}
\end{equation}
Each ``mode" $n$ in the expansion (\ref{mode}) gives rise to a corresponding one-dimensional Schrodinger equation (\ref{modeq}) with potential $V_n\equiv \alpha_n/v^2$.  In order to calculate the wave function $\psi$ and the entanglement entropy, we solve equations of the form (\ref{modeq}) numerically with boundary conditions\footnote{\textcolor{black}{It is interesting to note that reduced Schr\"odinger equation (\ref{modeq}) can be solved numerically in terms of Bessel functions after introducing a mode decomposition of the form $\xi_n(t,v) = \sum_j e^{-iE_{nj}t}\zeta_{nj}(v)$.  As discussed in detail in \cite{kunstatter2009polymer}, it turns out that the boundary condition $\lim_{v\to 0}\xi_n = 0$ is not enough to uniquely fix the mode solutions $\zeta_{nj}(v)$, something which results in a one-parameter family of self-adjoint extensions of the Hamiltonian.  However, this ambiguity plays no role in the numeric solution of (\ref{modeq}) since all self-adjoint extensions have $\lim_{v\to 0}\xi_n = 0$.  For a more comprehensive discussion of how various self-adjoint extensions modify the analytic mode solution of (\ref{tdse}) in the context of a mode expansion, interested readers are directed to \cite{Gielen:2020abd}.}}
\begin{equation}
    \lim_{v\rightarrow 0}\xi_n(t,v)= \lim_{v\rightarrow \infty}\xi_n(t,v)=0.
\end{equation}
To study entanglement dynamics with respect to the dust time $t$, we consider initial product states for the total wavefunction at $t=0$ of the form 
\begin{equation}
	\psi(0,v,\varphi) = f(v)g(\varphi) = f(v) \left[ \frac{1}{\sqrt{R}}\sum_{n=-\infty}^\infty c_{n} \exp \left( i \frac{2n\pi \varphi}{R} \right) \right].
\end{equation}
This means we are interested in solutions of the reduced Schrodinger equation (\ref{modeq}) with initial data $\xi_n(0,v)=f(v)$; we refer to solutions of (\ref{modeq}) with this initial data as $\xi_n^f(t,v)$. We select $f(v)$ to be peaked at some initial volume $v_0$ with momentum $p_0$:
\begin{equation}
	 f(v) = \mathcal{N} \exp\left[ -\frac{(v-v_{0})^{2}}{4\sigma^{2}} + ip_{0}v \right].
  \label{data1}
\end{equation}
The width $\sigma$ of these initial states is chosen small enough to ensure that the boundary condition $\lim_{v\to 0}\psi=0$ holds to within numerical accuracy at the initial time. For $p_0$ negative, the universe is initially hurtling toward the would-be singularity. The reduced density matrix for the evolution of such initial states, traced over the scalar field $\varphi$, is  
\bea
\rho(t, v, v') = \sum_{n=-\infty}^\infty |c_n|^2\ \xi^f_n(t,v)\ \xi^{f*}_n(t,v').
\eea}

\begin{figure*}
     \begin{center}
     \includegraphics[trim={150 150 200 100},clip,width=0.48\columnwidth]{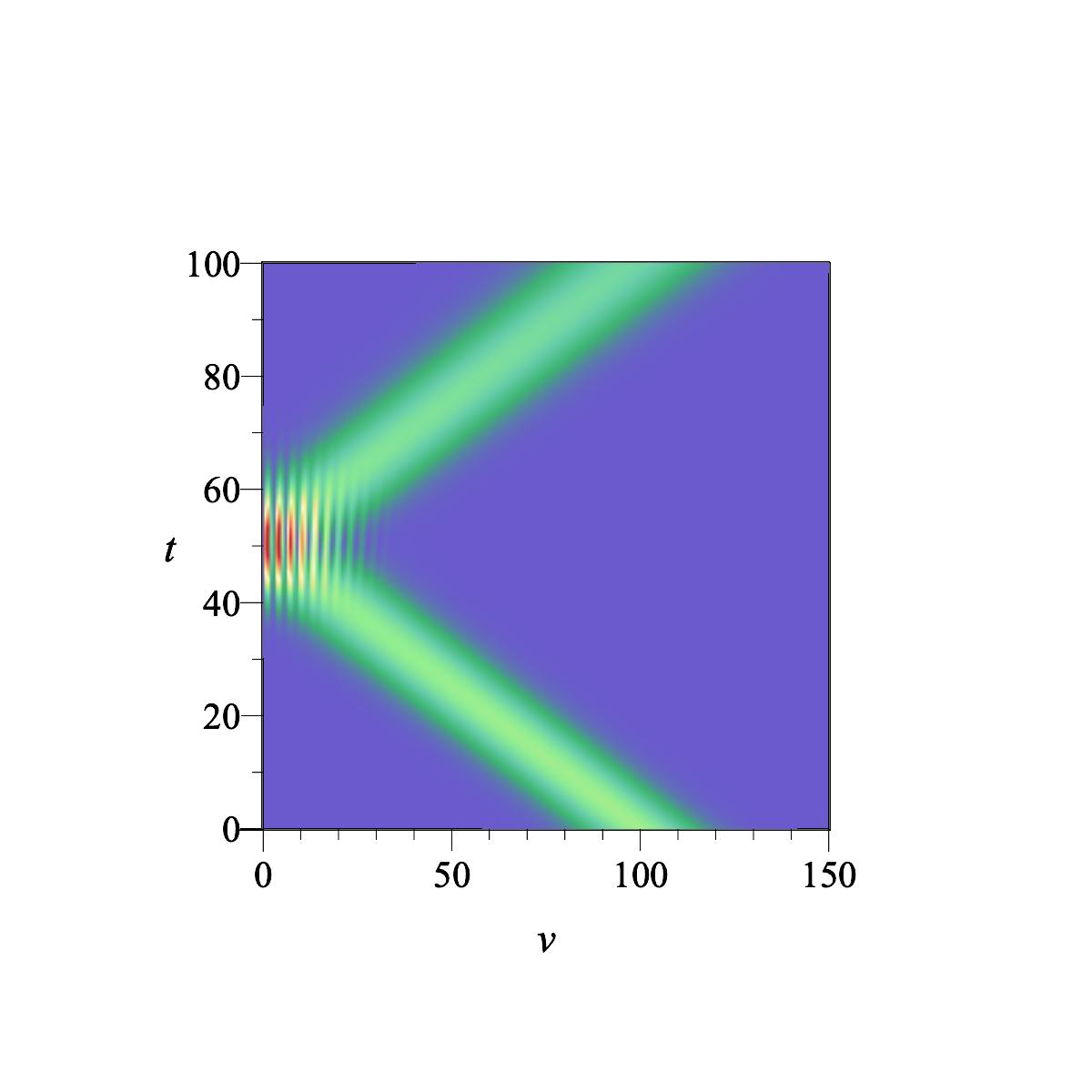}
     \includegraphics[trim={175 10 375 75},clip,width=0.51\columnwidth]{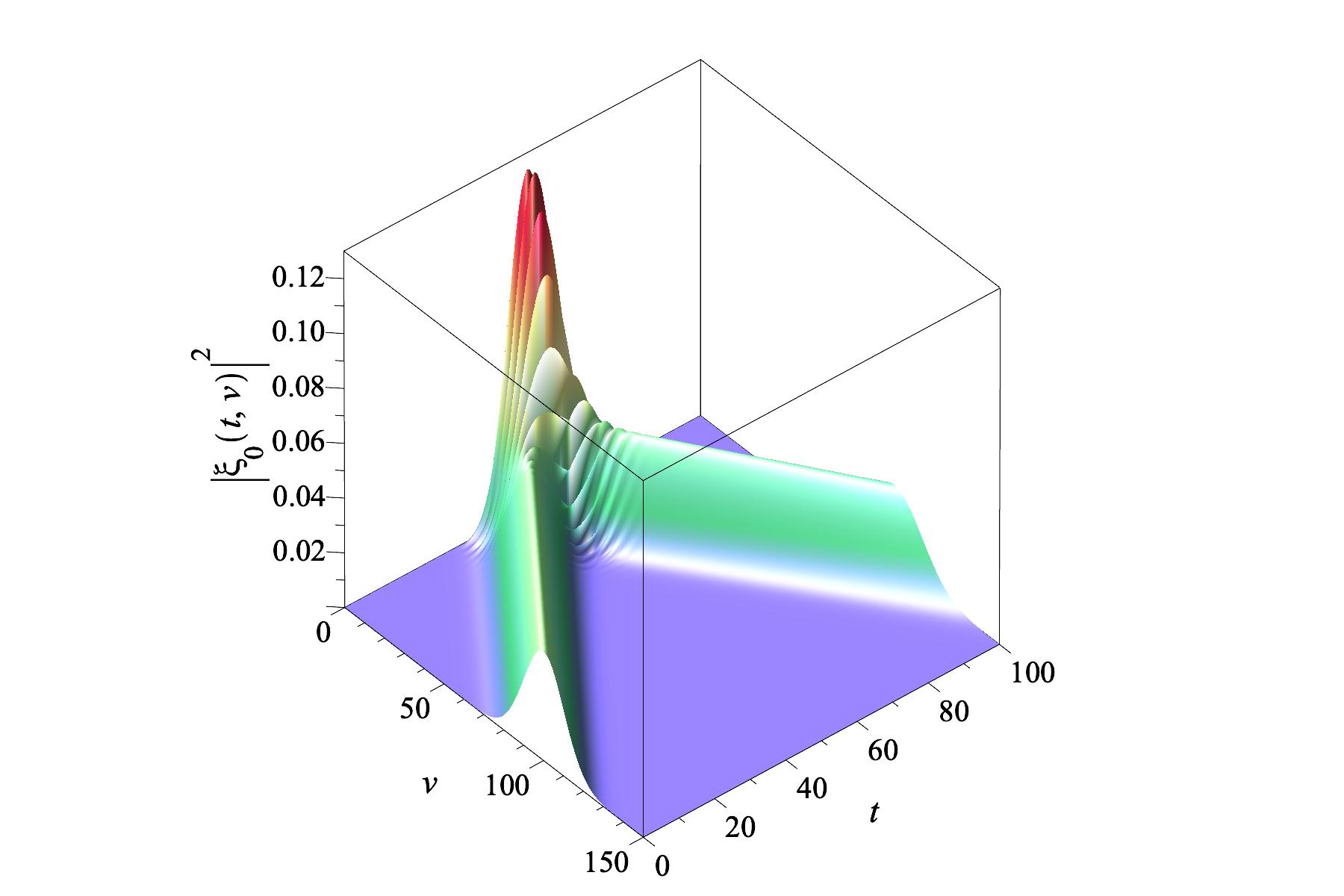}
    \end{center}
    \caption{A cosmological bounce showing the evolution of the volume probability density of an initial Gaussian wave packet  (\ref{data2}) with $\mu =1/4$, $R = 10$, $v_0=100$, $\sigma=5$, $p_{0} = -1$, $\epsilon=0$, and $m=0$. The bounce is at $t\approx50$ (in Planck units); the final density spread is larger than the initial one, indicating an asymmetric bounce.}
    \label{fig1}
\end{figure*}

It is possible to numerically compute the solutions  $\xi^f_n(t,v)$, the reduced density matrix, and one of the measures of quantum entanglement---the von Neumann entanglement entropy $ S_{\text vN}(t) =-\sum_i \lambda_i(t)\ln \lambda_i(t)$---from its eigenvalues $\lambda_i(t)$.  The specific initial states we consider are the ``two-mode" product states
\bea
\psi_{mn}(0,v,\varphi) = f(v)\left(\sqrt{1-\epsilon}\ e^{2\pi i m \varphi/R}+ \sqrt{\epsilon}\ e^{2\pi i n \varphi/R} \right)
\label{data2}
\eea
for various choices of $\epsilon\in [0,1]$ and $(m,n)$. Such states evolve to entangled states of matter and gravity. Fig. \ref{fig1} shows a typical evolution for $\mu =1/4$, $R = 10$, $v_0=100$, $\sigma=5$, $p_{0} = -1$, $\epsilon=0$, and $m=0$; the universe's contraction to the bounce at $t\approx50$ (in Planck units) and subsequent expansion is evident in the two frames (the first is a top-down view of the second). 

\begin{figure*}
    \begin{center}
        \includegraphics[width=0.32\textwidth]{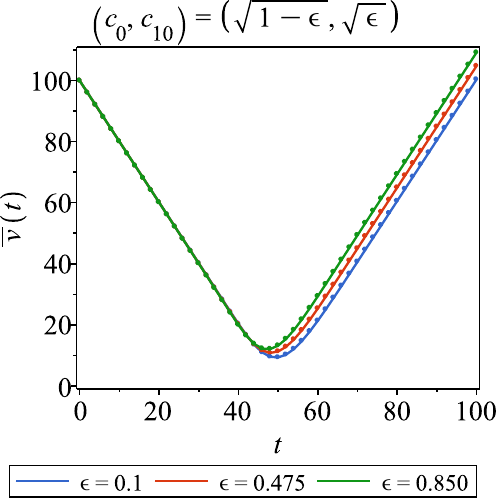}
        \hfill
        \includegraphics[width=0.32\textwidth]{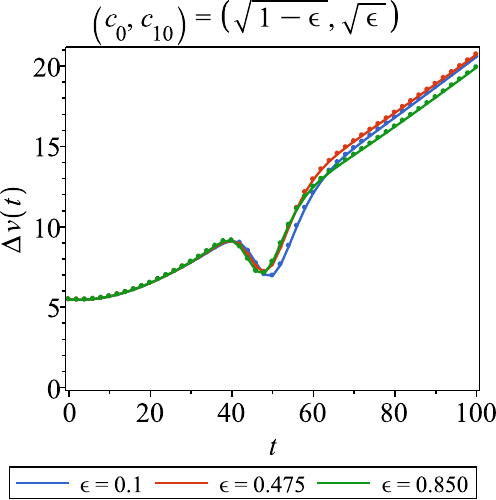}
   \hfill
        \includegraphics[width=0.32\textwidth]{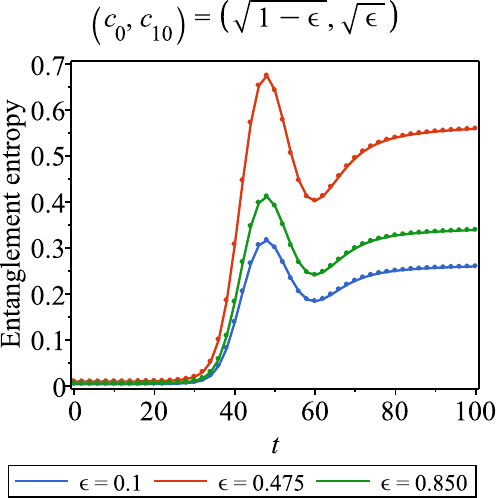}
   \hfill
    \end{center}
    \caption{Details of a cosmological bounce: the average volume, its variance, and matter-gravity entanglement entropy as the universe undergoes a bounce at $t\approx 50$ (in Planck units) for the initial state (\ref{data2}) with the parameters indicated. Notable features are the dips in the variance, the rapid rise and fall of the entropy, and its final saturation value (the other parameters are the same as those for Fig. \ref{fig1}).}
    
    \label{fig2}
\end{figure*}

Fig. \ref{fig2} illustrates the dust-time evolution of the expectation value of the volume $\bar{v} = \displaystyle \langle \hat{v}\rangle$, its variance $\displaystyle \Delta v  = \sqrt{\left\langle (\hat{v}-\bar{v})^2\right\rangle}$, and the von Neumann entropy as the state evolves. It is evident from the graphs of $\bar{v}$ and $\Delta v$ that evolution through the bounce is not symmetric; the wave function starts to disperse, narrows through the bounce, and continues to spread thereafter; its final width is significantly wider than the initial one over the symmetric time interval illustrated in Fig. \ref{fig2}.  Thus, although not noted in some earlier works where dispersion dynamics was not studied (e.g., \cite{Agullo:2016tjh}), our results indicate that asymmetric bounces are a natural feature in quantum cosmology, due entirely to the quantum dispersion of wave packets.

Our main result is the evolution of entanglement entropy through the bounce: its rapid rise through the bounce quantitatively captures the intuition of strong coupling between matter and gravity in regions of high curvature, and its subsequent relaxation to a constant nonzero value (noted earlier in \cite{Husain:2019nym}) indicates that matter-gravity entanglement may exist in our universe at present. These features are a consequence of the $p^2_{\varphi}/v^2$ coupling in the Hamiltonian (\ref{H}), which indicates large coupling at small volume, and hence  large entanglement. The qualitative features of the graphs in Fig. \ref{fig2} are similar for other values of $(m,n)$, and the other parameters, indicating the general robustness of the entropy rise and decline through the bounce and the subsequent saturation. It is also noteworthy that these features of our results do not depend on the choice of self-adjoint extension of the Hamiltonian. Rather, they follow directly from the form of the Hamiltonian. This is because  any initial product state must exhibit a rise in entanglement entropy from zero due to the increase in subsystem interaction as the volume decreases, followed by a freezing of the entropy as the volume increases after the bounce.

There is discussion in the literature of the possibility of a ``Second Law for entanglement entropy" despite the many differences from the familiar thermodynamics version of the Second Law. A recent theorem arising from considerations of a general form of entanglement manipulation of quantum states yields the result that there is no Second Law of entanglement entropy \cite{Lami_2023}. In this context, our calculation supports this theorem, and may be interpreted as an explicit form of entanglement manipulation coming from the cosmological Hamiltonian (\ref{H}); gravity and matter subsystems  are less interacting due to the $1/v^2$ ``coupling" at large volume, and more interacting at small volume.

Lastly, calculations similar to what we have carried out are possible in the loop quantum gravity framework applied to cosmology \cite{bodendorfer2016elementary,Agullo:2016tjh}, in particular in the dust time gauge \cite{Husain:2011tm}; it will be surprising if the main qualitative features of our results, the asymmetric bounce due to dispersion and the rise and decline of entanglement entropy through the bounce, differ in any significant way, since that framework may be viewed as a specific lattice version of the Schrodinger equation (\ref{tdse}).

\vskip 0.5cm
\noindent{\bf Acknowledgements:} This work was supported in part  by the Natural Science and Engineering Research Council of Canada.

\vskip 1cm

\bibliography{essay}

\end{document}